\documentstyle[sprocl]{article}

\input{psfig}
\def\Journal#1#2#3#4{{#1} {\bf #2}, #3 (#4)}

\def\NPB{{\em Nucl. Phys.} B}
\def\PLB{{\em Phys. Lett.}  B}

\def\beq{\begin{equation}}
\def\eeq{\end{equation}}
\def\bea{\begin{eqnarray}}
\def\eea{\end{eqnarray}}

\def\L{\Lambda}
\newcommand{\Tr}{\mathop{\mathrm{Tr}}}

\begin{document}

\title{SYMMETRY NON-RESTORATION AT HIGH \\ TEMPERATURE
\footnote{Talk given at the International Workshop {\em Beyond the 
Standard Model: from Theory to Experiment}, October 13-17, 1997, 
Valencia (Spain). It is based on work done with M.B. Gavela, O. P\`ene, 
M. Pietroni, N. Tetradis and S. Vargas-Castrill\'on
in refs. \cite{mnn,bons}.}}

\author{N. RIUS}

\address{Depto. de F\'\i sica Te\'orica and IFIC, Centro Mixto \\
 Universidad de Valencia-CSIC, Valencia, Spain}


\maketitle\abstracts{We discuss the (non)-restoration of 
global and local symmetries at high temperature. 
First, we analyze a two-scalar model with $Z_2 \times Z_2$ 
symmetry using the exact renormalization group. We conclude that 
inverse symmetry breaking is possible 
in this kind of models within the perturbative regime.
Regarding local symmetries, 
we consider the $SU(2) \otimes U(1)$ gauge symmetry 
and focus on the case of a strongly interacting scalar sector. 
Employing a model-independent chiral Lagrangian 
we find indications of symmetry restoration.}
  
\begin{flushright} 
FTUV/98-5 \\ IFIC/98-5
\end{flushright}

\section{Review of Symmetry Non-restoration and Inverse Symmetry 
\\ Breaking}

It has been known for a long time \cite{Weinberg}, 
that simple multi-scalar models can exhibit
an anti-intuitive behaviour associated with more broken symmetry as the 
temperature is increased. We refer to this behaviour as 
inverse symmetry breaking or symmetry non-restoration,  
depending on whether the symmetry is exact or not at zero temperature.
This phenomenon may have remarkable consequences for cosmology 
\cite{Dvali}. 

In renormalizable supersymmetric theories 
internal symmetries are always restored at high temperature, 
and there are arguments of why this should be also the case even when 
non-renormalizable terms are considered; 
symmetry non-restoration could be possible though 
in the presence of non-vanishing background charges
\cite{susy}.

Doubts have been raised on the validity of these results, since they 
are based on the one-loop approximation to the finite temperature 
effective potential, 
which is known to be unreliable for the discussion
of many aspects of phase transitions.
Different techniques are being actively applied to improve the 
one-loop approximation, leading quite often to contradictory results.
The effect of next-to-leading-order contributions within
perturbation theory has been investigated  
\cite{Bimonte1}, through
the study of gap equations, which are
equivalent to a resummation of the super-daisy diagrams of the
perturbative series. Large subleading corrections have been identified, 
which lead to a sizeable reduction of the parameter space where inverse 
symmetry breaking occurs. The question of inverse symmetry breaking
has also been studied through the use of the renormalization group
\cite{Roos,mnn} and a variational approach \cite{Camelia}, 
with similar conclusions.  
Contrary to the results of the above studies, 
a large-$N$ analysis seems to indicate
that symmetry is always restored at high temperature  \cite{Fujimoto}.
However, the validity of this claim has recently been questioned  
\cite{Orloff}. A finite-lattice calculation also supports symmetry
restoration at sufficiently high temperature \cite{Bimonte2}. 
Although the relevance of this result for the continuum limit is 
not clear, a Monte Carlo simulation in 2+1 dimensions seems to 
support this conclusion \cite{Bimonte3}.

\section{Inverse Symmetry Breaking and the Renormalization Group}

In this section 
we consider the simplest model that exhibits inverse symmetry breaking:
a two-scalar model with $Z_2 \times Z_2$ symmetry. 
The tree-level potential is given by 
\beq
V_{tr}(\phi_1, \phi_2) = \frac{1}{2} m_1^2 \phi_1^2 +\frac{1}{2} m_2^2 \phi_2^2
+ \frac{1}{4} \lambda_1 \phi_1^4 + \frac{1}{4} \lambda_2 \phi_2^4 -
\frac{1}{2}\lambda_{12} \phi_1^2 \phi_2^2~.
\label{pot}
\eeq
This potential is bounded for $\lambda_{1,2}>0$ 
and 
\beq
\lambda_1 \lambda_2 > \lambda_{12}^2~.
\label{stab} 
\eeq
In the high temperature limit, $|m_i| \ll T$, 
the thermal correction to the above potential at the one-loop level 
is given by \cite{Dolan} 
\beq
\Delta V_{T}(\phi_1, \phi_2) \simeq 
\frac{T^2}{24}\left[(3\lambda_1 - 
\lambda_{12}) \phi_1^2 + (3 \lambda_2 - \lambda_{12}) \phi_2^2 \right]~+~\dots
\label{pert}
\eeq
For the parameter range 
\beq
3 \lambda_1 - \lambda_{12} < 0~,
\label{pertpred}
\eeq
which can be consistent with the stability condition of eq. (\ref{stab}), 
the thermal correction for the mass term of the $\phi_1$ field is negative. 
Notice that the stability condition (\ref{stab}) does not allow 
both mass terms to be negative.
If the system is in the symmetric phase at zero temperature with
$m_{1,2}^2>0$, there will be a critical temperature 
$T_{cr}^2 = 12 m_1^2/(\lambda_{12}
-3\lambda_1)$ above which the symmetry will be broken.
If the system is in the broken phase at $T=0$, the symmetry will never be 
restored by thermal corrections. 

Our aim is to discuss the above scenario
in the context of the Wilson approach to the renormalization group. 
The main ingredient in this approach is an exact flow equation that 
describes how the effective action of the system evolves
as the ultraviolet cutoff is lowered. We consider the lowest order in a
derivative expansion of the effective action, which contains 
a general effective potential and a standard kinetic term.
At non-zero temperature this approach can be formulated either in the 
imaginary-time \cite{Nick1} or in the real-time formalism \cite{Max}. 
In the latter formulation, the evolution of the potential lowering 
the cutoff scale $\Lambda$ is  given by 
the partial differential equation
\cite{Max}
\beq
\L\frac{\partial\:}{\partial \L} 
V_{\L}(\phi_1, \phi_2) = -T\frac{\L^3}{2 \pi^2}
\Tr\left\{ \log\left[1-\exp\left(-\frac{1}{T}\sqrt{\L^2+{\cal M}_\L^2}\right)
\right]\right\}~,
\label{4dr}
\eeq
where 
\beq
\left[{\cal M}_\L^2(\phi_1, \phi_2)\right]_{i,j} = 
\frac{\partial^2 V_{\L}(\phi_1, \phi_2)}{\partial \phi_i \partial \phi_j}~, 
\;\;\;\;\;\;\;\;i,j = 1,2 ~.
\label{massma}
\eeq
The initial condition for the above equation, at 
a scale $\L_0 \gg T$, is the renormalized effective
potential at zero temperature.
We consider small quartic couplings,  so that the  
logarithmic corrections of the zero-temperature theory can be
safely neglected. The initial condition for the evolution is a 
zero-temperature potential given by eq. (\ref{pot}).  
Integrating the evolution equation (\ref{4dr}), we obtain the 
non-zero-temperature effective potential
in the limit $\L\rightarrow 0$.

Finding the solution of eq. (\ref{4dr}) is a difficult
task. An approximate solution can be obtained 
by expanding the potential in a power series in the fields. 
In this way the partial differential equation (\ref{4dr}) is transformed 
into an infinite
system of ordinary differential equations for the 
coefficients of the expansion. 
This system can be solved approximately by truncation at a finite 
number of equations. 
That is, 
the potential is approximated by a finite-order polynomial. 
As a first step, we follow this procedure and define   
the running masses and couplings at the origin 
\beq
m^2_{1,2}(\L)= \left. \frac{\partial^2 V_\L}{\partial \phi_{1, 2}^2}
\right|_{\phi_{1,2}=0},\:\:\:\:\:\:\:
\lambda_{1,2}(\L)= \left. \frac{1}{6}
\frac{\partial^4 V_\L}{\partial \phi_{1, 2}^4}
\right|_{\phi_{1,2}=0},
\:\:\:\:\:\:
\lambda_{12}(\L)= - \left. 
\frac{1}{2}\frac{\partial^4 V_\L}{\partial \phi_{1}^2
\partial \phi_{2}^2}
\right|_{\phi_{1,2}=0}~.
\label{param}
\eeq
The corresponding evolution equations can be obtained by 
differentiating eq. (\ref{4dr}) and neglecting the higher derivatives
of the potential. We find 
\bea\label{trunc}
 \L \frac{\partial \:}{\partial\L} m_{1,2}^2 &=& - 6 C_{1,2} 
\lambda_{1,2} 
+ 2 C_{2,1} \lambda_{12} \nonumber \\
\L \frac{\partial \:}{\partial\L} \lambda_{1,2} &=& - 18 D_{1,2}
 \lambda_{1,2}^2 
- 2 D_{2,1} \lambda_{12}^2  \\
\L \frac{\partial \:}{\partial\L} \lambda_{12} &=& - 6 D_1
 \lambda_{1} \lambda_{12}  - 6 D_2 \lambda_{2} \lambda_{12}
  +8 \frac{C_1 - C_2}{{m_1}^2 - m_{2}^2} \lambda_{12}^2~,\nonumber
\eea
with
\beq
C_{1, 2}=\frac{\L^3}{4 \pi^2}\frac{N(\omega_{1,2})}{\omega_{1,2}}~, 
\:\:\:\:\:\:\:\:
D_{1, 2} = \frac{\partial 
C_{1,2}}{\partial m_{1,2}^2}~,\:\:\:\:\:\:\:\: 
\omega_{1,2}^2=\L^2 + m_{1,2}^2~,
\label{def1} 
\eeq
and $N(\omega)=\left[ 
\exp(\omega/T)-1 \right]^{-1}$ the Bose-Einstein distribution 
function.
For $\omega_{1,2} \ll T$ we have
\beq
C_{1,2}\rightarrow \frac{\L^3}{4\pi^2}\frac{T}{\L^2+m_{1,2}^2}~,
\label{limit} \eeq
and the above equations agree with those considered in ref. \cite{Roos}
in the same limit.
For $\omega_{1,2} \gg T$ there is no 
running, because of 
the exponential suppression in the Bose-Einstein function. 

We have solved numerically the system of equations (\ref{trunc}) 
and determined the range of zero-temperature parameters
that lead to inverse symmetry breaking. 
In fig.~1 we present the results for a zero-temperature theory
with positive mass terms $m^2_1(\L_0)=m_2^2(\L_0)$ and $\lambda_2(\L_0)=0.3$. 
The temperature has been chosen much higher than the 
critical one (\mbox{$T=500 m_1(\L_0)$}).  
The system  (\ref{trunc}) has been integrated from $\Lambda_0 \gg T$ down to
\mbox{$\L = 0$}, where the thermally corrected masses and 
couplings at non-zero temperature have been obtained. 
A negative value for the mass term $m^2_1$ at $\L=0$
has been considered as 
the signal of inverse symmetry breaking. This has been achieved 
in the region above line (a) in fig.~1. 
We also plot the stability bound of eq. (\ref{stab}) 
(the allowed range is below line (b)), and  
the perturbative prediction for the range that 
leads to inverse symmetry breaking (above line (c)).  
The phenomenon of inverse symmetry breaking is confirmed by our study,
in agreement with ref. \cite{Roos}, where the imaginary-time 
formulation of the renormalization-group approach has been used. 
We observe that the renormalization-group treatment eliminates 
a large part of the parameter space allowed by perturbative theory,
in agreement with the results obtained by solving 
the gap-equations \cite{Bimonte1}.

\begin{figure}
\psfig{figure=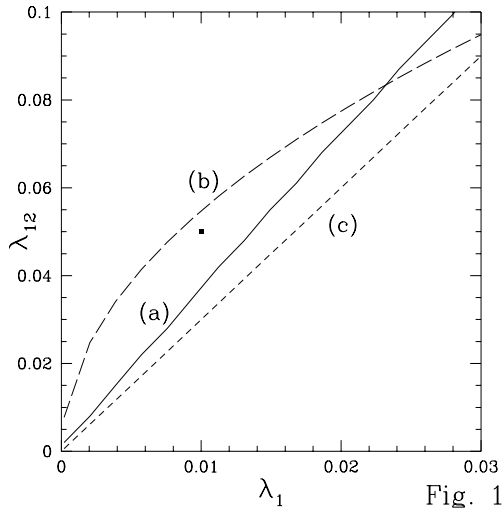,rwidth=2.5in,rheight=2.5in,width=2in,height=3in,bbllx=-160pt,bblly=-30pt,bburx=280pt,bbury=600pt}
\caption{Parameter space that leads to inverse symmetry breaking 
($\lambda_2 = 0.3$).}
\end{figure}

The reliability of our conclusions crucially depends  on whether 
the solution of the system of truncated equations 
(\ref{trunc}) provides an approximate solution to the 
full partial differential equation (\ref{4dr}). 
We have checked that by 
numerical integration of eq. (\ref{4dr}) through the algorithms 
discussed in ref.~\cite{Adams}. 
Due to limitations in computer time, 
we restrict our discussion of eq. (\ref{4dr})
along the $\phi_1$ axis, which is
the direction of expected symmetry breaking for our choice of couplings. 
We approximate the potential by the expression
\beq
V_\L(\phi_1, \phi_2) = 
V_\L(\phi_1) +\frac{1}{2} m_2^2(\L) \phi_2^2
+ \frac{1}{4} \lambda_2(\L) \phi_2^4 -
\frac{1}{2}\lambda_{12}(\L) \phi_1^2 \phi_2^2~.
\label{potphi1}
\eeq
The evolution of 
$m_2^2(\L)$, $\lambda_2(\L)$ and 
$\lambda_{12}(\L)$ is determined through the truncated eqs.
(\ref{trunc}). However, the full $\phi_1$ dependence is preserved through
the numerical integration of eq. (\ref{4dr}), with the eigenvalues of the
mass matrix ${\cal M}_\L^2$ given by 
\beq
\left[ {\cal M}_\L^2 \right]_1 = 
\frac{\partial^2 V_\L(\phi_1)}{\partial \phi^2_1}
~~~~{\rm and}~~~~~~
\left[ {\cal M}_\L^2 \right]_2 = 
m_2^2(\L) - \lambda_{12}(\L) \phi_1^2. 
\label{eigen}
\eeq
This treatment permits a reliable study of the order of the 
symmetry-breaking phase transition; we have found that it is 
governed by the Wilson-Fisher fixed point of the one-scalar 
three-dimensional theory, resulting in a second order phase 
transition \cite{mnn}.

\section{$SU(2) \otimes U(1)$ gauge symmetry with strongly coupled Higgs
sector}

A natural question to ask is whether the standard model gauge symmetry
$SU(2) \otimes U(1)$ could remain broken at high temperatures. 
It is well known that the symmetry is restored in the 
minimal standard model, so we consider its simplest extension with 
two Higgs doublets. We find that 
due to the positive and large contribution of the gauge bosons to 
the scalar thermal masses, it is not possible to attain a
negative mass term (needed for symmetry non-restoration)
within the perturbative range of the scalar couplings.

A strongly coupled Higgs sector implies (at least naively) heavy 
physical scalar particles, which can be effectively removed 
from the physical low-energy spectrum. 
To study the behaviour of the gauge symmetry in this case
we use an effective Lagrangian which keeps only the light degrees of 
freedom, namely the gauge and Goldstone bosons together with 
the fermions.
The resulting chiral Lagrangian is a non-renormalizable non-linear 
sigma model coupled in a gauge invariant way to the Yang-Mills 
theory \cite{al}. At lowest order, it is model independent.

We propose to use the gauge boson magnetic mass  
as an indicator of symmetry (non)-restoration. It is
defined as the transverse part of the corresponding 
self-energy, $\Pi_T(0, \vec{k})$, on-shell.
A perturbative computation shows that it is exactly 
equal to zero at one loop in an unbroken gauge theory. 
For unbroken non-Abelian gauge theories, such as QCD, higher orders in 
perturbation theory suffer from infrared divergences, and a magnetic 
mass of order $g^2 T$ is expected to be generated 
non-perturbatively. In spontaneously broken gauge theories, such as the
standard electroweak model and its extensions, no such divergences 
are present. 
Thus, we expect that even in perturbation theory the magnetic mass 
(as computed in the broken phase) will show 
a tendency to vanish at high enough temperatures, 
whenever symmetry restoration occurs. 

We have calculated the thermal gauge boson self-energies at one 
loop and leading order, ${\cal O} (T^2)$, from which we obtain 
the magnetic masses  
\bea
M_{W,mag}^{2} &=&  g^2 \, \frac{v(T)^2}{4} \ ,
\\
M_{Z,mag}^{2} &=&  (g^2 + g'^2) \, \frac{v(T)^2}{4} \ ,
\eea
with 
\beq
v(T)^2 = v^{2}
\left[ 1 - \frac{T^2}{6 v^{2}} \right]  \ ,
\label{vev}
\eeq
where the fact that the gauge couplings are not renormalized at this 
order has been used. 
We conclude thus that in models with strongly 
interacting Higgs sector the spontaneously broken 
$SU(2) \otimes U(1)$ gauge symmetry 
tends to be restored when the system is heated. 
Notice though that our calculation is only valid for temperatures below
the electroweak scale.
Details of the calculation and the approximations involved can be 
found in \cite{bons}.

It is worth to remark that the thermal corrections to 
$v$ coincide with those of the pion decay constant $F_\pi$
in the non-linear sigma model \cite{fpi}.  
That is, at one loop and leading order $T^2$
all the temperature corrections to the scalar v.e.v. are due to 
the would-be Goldstone bosons; fermions and gauge boson transverse 
degrees of freedom will only contribute at higher order.

\section*{Acknowledgements}

It is a pleasure to thank all my collaborators in the study of 
symmetry non-restoration: M.B. Gavela, O. P\`ene, M. Pietroni, 
N. Tetradis and S. Vargas-Castrill\'on.
This work was supported in part by CICYT under grant AEN96-1718,
by DGICYT under grant PB95-1077 (Spain) and by EEC under the 
TMR contract ERBFMRX-CT96-0090.

\end{document}